\documentclass[12pt]{article}
\usepackage{cite}
\usepackage{amsmath, amsthm, amssymb,slashed,graphicx,epsfig}

\newcommand{\smallfrac}[2] {\mbox{$\frac{#1}{#2}$}}
\newcommand {\slsh} [1] {\not{\hbox{\kern-2pt${#1}$}}}

\newcommand{\gsim}{\lower.7ex\hbox{$\;\stackrel{\textstyle>}{\sim}\;$}}
\newcommand{\lsim}{\lower.7ex\hbox{$\;\stackrel{\textstyle<}{\sim}\;$}}
\newcommand {\beq} {\begin{equation}}
\newcommand {\eeq} {\end{equation}}
\newcommand {\beqn}{\begin{eqnarray}}
\newcommand {\eeqn} {\end{eqnarray}}
\newcommand{\bea}{\begin{eqnarray}}
\newcommand{\eea}{\end{eqnarray}}

\def\Z{\mathbb Z}
\def\1{\mathbbm{1}}

\def\1{\mbox{\tiny (1) }}
\def\0{\mbox{\tiny (0) }}

\def\be{\begin{eqnarray}}
\def\ee{\end{eqnarray}}

\def\nablaslash{\,\,{\raise.15ex\hbox{/}\mkern-12mu \nabla}}
\def\nablabarslash{\,\,{\raise.15ex\hbox{/}\mkern-12mu {\bar \nabla}}}
\def\Dslash{\,\,{\raise.15ex\hbox{/}\mkern-12mu D}}
\def\Dbarslash{\,\,{\raise.15ex\hbox{/}\mkern-12mu {\bar D}}}
\def\delslash{\,\,{\raise.15ex\hbox{/}\mkern-9mu \partial}}
\def\delbarslash{\,\,{\raise.15ex\hbox{/}\mkern-9mu {\bar\partial}}}
\def\pslash{\,\,{\raise.15ex\hbox{/}\mkern-9mu p}}
\def\calDslash{\,\,{\raise.15ex\hbox{/}\mkern-12mu {\cal D}}}

\def\lae{\mathrel{\mathop{\smash{\lower .5 ex \hbox{$\stackrel<\sim$}}}}}
\def\lae{\mathrel{\mathop{\smash{\lower .5 ex \hbox{$\stackrel>\sim$}}}}}

\def\beqn{\begin{eqnarray}}
\def\eeqn{\end{eqnarray}}

\def\ba{\beq\new\begin{array}{c}}
\def\ea{\end{array}\eeq}
\def\be{\ba}
\def\ee{\ea}

\begin{document}

\begin{titlepage}

\begin{flushright}
FTPI-MINN-09/17; UMN-TH-2745/09\\
May 2009
\end{flushright}

\vskip 0.4in
\begin{center}
{\bf\Large{Quantum Fusion of Strings (Flux Tubes) 
\\[2mm]
and Domain 
Walls}}\vskip0cm
\vskip 0.5cm  {\bf\large{S.~Bolognesi$^a$,  M.~Shifman$^a$,  and  M.B.~Voloshin$^{a,b}$}} \vskip
0.05in $^a${\small{ \textit{William I. Fine Theoretical Physics Institute, University of Minnesota, } \\ {\textit{116 Church St. S.E., Minneapolis, MN 55455, USA}}}}
\vskip
0.05in $^b${\small{ \textit{Institute of Theoretical and Experimental Physics,117218, Moscow, Russia } }}
\end{center}
\vskip 0.5in

\baselineskip 10pt
%

\begin{abstract}
We consider formation of composite strings and domain walls as a result of fusion of two elementary objects
(elementary strings in the first case and elementary walls in the second) located at a distance from each other. 
The tension of the composite object $T_2$ is assumed to be less than twice the tension of the elementary object
$T_1$, so that bound 
states are possible. If in the initial state the distance $d$ between the fusing strings or walls
is much larger than their thickness and satisfies the conditions $T_1d^2 \gg 1$ (in the string case)
and $T_1d^3 \gg 1$ (in the wall case), the problem can be fully solved quasiclassically.
The fusion probability is determined by the first, ``under the barrier"  stage of the process.
We find the bounce configuration and its extremal action $S_B$. In the wall problem $e^{-S_B}$
gives the fusion probability per unit time per unit area. In the string case, due to
a logarithmic infrared divergence, the problem is well formulated only for finite-length strings.
The fusion probability per unit time can be found in the limit in which the string length
is much larger than the distance between two merging strings.

\end{abstract}


\end{titlepage}
\vfill\eject


\section{Introduction}
\label{intro}

In this paper we will consider two problems of practical interest which arise in various settings, and can be solved purely quasiclassically. 
The formulation of the problems, and their  solution, is very general. They refer 
to restructuring of solitonic objects (or branes) supported in various  field  theories. 
We were motivated by a specific problem that arose
in \cite{one}, but here we will give a general discussion, and find a generic solution, 
so that our results can be used in all similar situations.

The first problem is about strings (flux tubes). 
Suppose we have two types of strings: ``elementary'' strings with tension $T_1$, and a composite string with tension $T_2$. 
We assume that the composite string is a bound state, i.e.
\beq
T_2 -2 T_1 < 0 \ .
\eeq
By composite we mean that there is a conserved ``charge" $Q$, and $Q=1$ for the 
elementary string while $Q=2$ for the composite one.
The composite string can form as a result of a fusion of two elementary ones.

\begin{figure}[h!t]
\epsfxsize=9cm
\centerline{\epsfbox{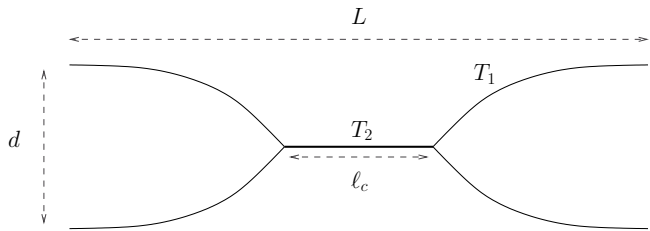}}
\caption{{\footnotesize Two elementary strings merge into a composite one. Once the size $\ell$ of the merged segment reaches (in a quantum tunneling process) its critical value $\ell_c$, further expansion of its size becomes classical. }}
\label{idea}
\end{figure}

One can consider two parallel strings at a distance $d$ from each other. The parameter $d$ is assumed to be much larger 
than the string thickness.
Quantum fluctuations of strings can result in a configuration with two elementary strings forming a composite one in the middle (see Fig.~\ref{idea}).
A crucial characteristic of the process is a critical size $\ell_c$ of the merged
segment.  It is determined by the balance of two energies: the one  gained due to the fact that $T_2 -2 T_1 < 0 $
(the energy gain is $\ell_c \left(2 T_1-T_2\right)$), and the one  lost due to elongation of the 
strings\,\footnote{In the present work it is assumed that the three-string junction mass is negligible.}.
The first stage of dynamics producing a size $\ell_c$
merged segment occurs as a quantum tunneling,\footnote{This is due to the fact that at $\ell<\ell_c$
the energy gain is less than the energy loss; therefore the system 
under consideration tunnels  under a barrier.}  which can be described in the Euclidean space-time.
Once the critical-size segment is attained, its further expansion proceeds as a purely classical process, 
with  positive energy release and an accelerating expansion of the merged segment.
The fusion probability is determined by the quantum tunneling stage.

The action corresponding to such fusion is large, provided
\beq
T_1\,  d^2 \gg 1 \ .
\label{ishco}
\eeq
As was mentioned, we gain energy in the central domain because $T_2 < 2 T_1$. 
We loose outside the central domain because of the string bending needed to match the 
asymptotic boundary conditions.
An extremal (in fact, maximal) value of the action must exist. It is realized on a classical solution
in Euclidean.
This is a maximum with respect to the size variation $\ell$. 
This is the only instability, and is the usual one that gives the vacuum decay.

The problem is similar, in a sense,  to that of
metastable vacuum decay \cite{two,three} (for reviews see
\cite{four,five}), but with an important difference (see Appendix~A). 
In the false vacuum decay the energy balance is achieved between 
a bubble of a genuine vacuum (gain) versus the potential energy of its surface (loss).
The Euclidean solution is provided by a bounce configuration. In the problem at hand, in which 
the three-string
junction is assumed to carry no energy, the
barrier is not due to the potential energy, but is rather associated with the kinetic energy term in the
string  Lagrangian. However, this is just a technical difference. A critical field configuration extremizing
the Euclidean action still exists, and we will find it in the limit when
the string thickness is negligible compared to the interstring distance $d$, see Fig.~\ref{idea}. 
This is an  analog of the standard bounce \cite{four}.\footnote{We will apply this term, bounce,
to the extremal Euclidean string and wall configurations in the problems to be discussed below.}

In the formulation of the string fusion 
problem there are infrared subtleties related to the  tails of the strings.
We can define the problem in a finite box, of length $L$, and compute, with exponential accuracy,
the probability 
per unit of time of this fusion process $\Gamma(L)$. This is the content of Section \ref{strings}.

The result of the above computation immensely simplifies  if one calculates the exponent
in a logarithmic approximation. In this approximation it turns out possible
(in the limit $2T_1-T_2 \ll T_1$) to generalize the analysis of the parallel string fusion
to cover the case of nonparallel strings. This problem will be addressed in Sect.~\ref{nonpar}.

The third problem we will deal with, in Section \ref{walls}, is similar in nature, but it refers to 
parallel domain walls, rather than strings. 
Adding an extra  dimension 
to the solitonic objects to be fused has  a crucial effect. The infrared problem we had to deal with
in the case of the string fusion now disappears, even in the
infinite volume. Then, we can readily calculate the fusion probability per unit time and per unit area, with the exponential accuracy. This is done in Sect.~\ref{walls}. In Sect.~\ref{pwlb} we deal with the wall merger at strong binding.
An instructive example of the domain wall fusion in super-Yang--Mills theory in considered in Sect.~\ref{aie}.

Is not difficult  to generalize our analysis to branes with arbitrary $p$ spatial dimensions, usually called $p$-branes, in a  
space-time with $D+1$ dimensions. We outline this, and summarize our results in Sect.~\ref{conq}.

In summary, our solution of the string/wall fusion problem is general and independent of dynamics of
the underlying
microscopic theory provided the following  assumptions are met:
(i) $\delta /d\to 0 $ where $\delta$ is the string or wall thickness; (ii) $T_1d^2$ (in the string case) or $T_1d^3$ (in the wall case) $\gg 1$;
(iii) the three-string (or three-wall) junction contribution to the extremal action is negligible. 
The latter condition is met in many instances of practical interest. In the string case we must also assume that
$L\gg d$. At weak binding the constraint on $d$ softens;
it is sufficient to require $T_1d^3\sqrt{\frac{T_1}{2T_1-T_2}}\, \gg 1.$

\section{Parallel Strings}
\label{strings}

To compute the decay probability, it is convenient to Wick-rotate the time direction. Then in the Euclidean space-time
we have a problem of two static
$2$-branes, which can fuse due to quantum fluctuations.\footnote{In Euclidean space one can view these
fluctuations as thermal.}
The Euclidean string action is  the string tension multiplied by the area of the branes.

\begin{figure}[h!t]
\epsfxsize=10cm
\centerline{\epsfbox{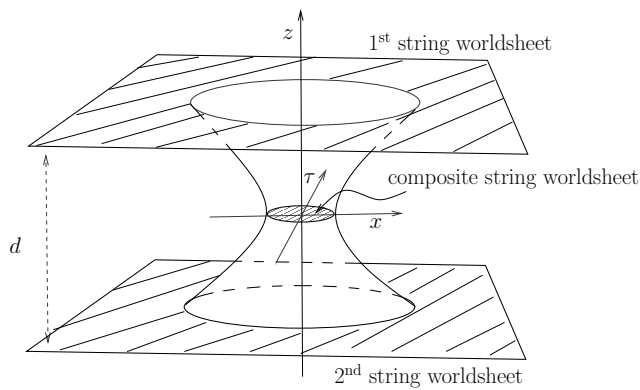}}
\caption{{\footnotesize World surface for two elementary strings forming a composite one. The Euclidean time is denoted by $\tau$. }}
\label{bouncestring}
\end{figure}

We want to find a bounce solution, which corresponds to two surfaces at asymptotic distance $d$, and an 
interior ``bubble" in which they overlap to form a bound state (Fig.~\ref{bouncestring}). The tunneling rate is then determined by the 
difference $S_B$ between the (Euclidean) action on the bounce configuration and 
that on the trivial configuration, with 
two flat world sheets for each string, one at $z= d/2$ and another at $z= -d/2$,
\beq
\Gamma = {\cal C} \, \exp (-S_B)~,
\label{gamsb}
\eeq
where ${\cal C}$ is a pre-exponential factor. 
Getting this factor requires calculation of the path integrals over 
fluctuations around the bounce solution, as well as around the trivial flat world-sheet
configurations. This issue will not be addressed here.
In what follows we will discuss only the exponential factor determined by 
classical solutions to the Euclidean equations of motion.

One can see, however, that for infinite strings a bounce solution does not exist. This is due to
an infrared peculiarity of two-dimensional surfaces. 
Whenever we pull such a surface in the perpendicular direction, it
never becomes flat asymptotically.
Its asymptotic behavior, from the solution to the Laplace equation, is always logarithmic. Thus, we have no chance to recover the required boundary condition,
that at $x,\,\tau \to\infty$ the $z$ coordinate of the surface tends to $\pm \, d/2$.

\begin{figure}[h!t]
\epsfxsize=8cm
\centerline{\epsfbox{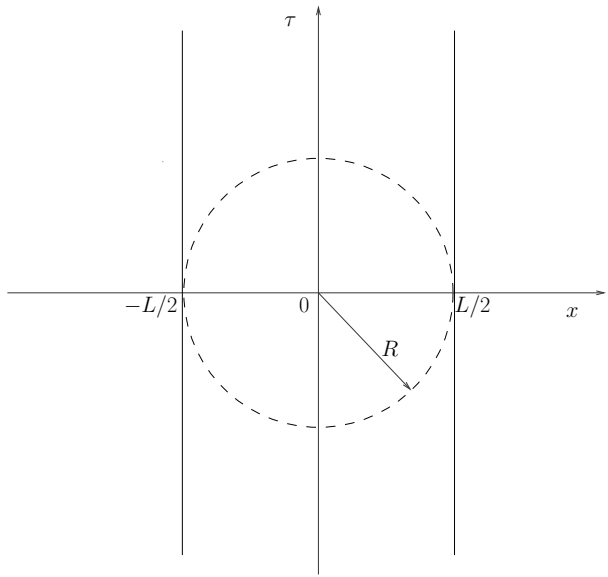}}
\caption{{\footnotesize  Boundary conditions for the fusing string worldsheet. Solid lines: ``strip'' boundary conditions. Dashed line: ``round'' boundary condition. }}
\label{boundary}
\end{figure}

This infrared behavior can be regularized provided we assume that $z =\pm d/2$
is achieved at some finite distance in the $\{x,\,\tau\}$ plane. The most physically transparent regularization
of the Euclidean version of the problem consists of a strip (infinite in the $\tau$ direction),
with the boundary conditions $z=\pm\,d/2$ implemented at its edges (Fig.~\ref{boundary}). 
We parametrize the coordinates as $x,\,\tau$ and $z$, 
where $x=\mp  L/2$ present two edges of the strip, $\tau$ 
corresponds to the Euclidean time, and $z=\pm d/2$ are the vertical locations 
of the two parallel strings, so that the boundary condition
for the bounce configuration sought for is as follows: at
$x=\pm L/2$ the value of $z$ is fixed at $+d/2$ for one string and at $-d/2$ for the other.
We use $x$ and $\tau$ to parametrize the brane, and $z= f (x,\tau)$ determines the height of the branes. 
In fact, it is sufficient to consider only the upper side of the picture since it is symmetric under reflection $z\to-z$.
This will be referred to as a ``strip" boundary condition.

Below we will find that $S_B$ depends on $L$ only through $\ln L$. Aiming at logarithmic accuracy
(i.e. keeping $\ln L$ and omitting nonlogarithmic constants assuming $\ln L$ to be large),
we can replace the strip boundary conditions by much simpler ones, to be referred to as
``round" boundary conditions (Fig.~\ref{boundary}). The round boundary condition
is convenient for two reasons. First, it will help us to calibrate our solution. Second, the 
results obtained with the round boundary condition are useful in
extending the problem to the case of nonparallel strings. 
The problem with the round boundary conditions is that, 
by itself, it has no Minkowski physical interpretation. In the 
Euclidean space, instead, it is just a problem of fusion due to thermal fluctuations, 
with the position of the $2$-branes fixed at the circle. 

\subsection{String fusion with round boundary conditions}
\label{round}

In this section we will require $z(R)=\pm\,d/2$ where \beq
R\gg d
\label{limitation}
\eeq
is assumed (for the definition of $R$ see Fig. \ref{boundary}).
In the leading logarithmic (in $R$) approximation the problem of merging of two parallel strings can 
be solved for arbitrary relation between the tensions $T_2$ and $T_1$ as long as the 
merger is possible, i.e. $T_2 < 2T_1$.
We do not have to require $2T_1-T_2\ll T_1$. Thus,  in this section we lift this constraint. 

Thus, we replaced the strip space-time 
boundary for the world sheets by 
a disk of a large radius $R \sim L/2$, so that 
a bounce centered at the origin (i.e. $x=0$ and $\tau=0$) is $O(2)$ axially symmetric and is described by 
a function $z(r)$ where 
\beq
r=\sqrt{x^2+\tau^2}\,. 
\eeq
The  slice of the solution  $z(r)$ passing through the $x=\tau=0$ line is 
shown in Fig.~\ref{idea} (where $\ell_c = 2r_c$ and $L$ must be replaced by $2R$). The 
contribution of each string's world sheet to the action for such a centrally symmetric configuration is given by the integral 
\beq
S=  2 \pi \, T \, \int r \, dr \, \sqrt{1+z^{\prime \, 2}} 
\eeq
with an  appropriate tension $T$. Here $z' = dz/dr$, and the integrand represents the area of
the circular element of the surface.

The central part of the bounce configuration, a disk of radius $r_c$ located at $z=0$, is filled by the string with tension $T_2$. The profile $z(r)$ for each of the strings with tension $T_1$ 
is determined by the equations of motion, which extremize the surface area of 
two world sheets. The difference $S_B$ of the action on the bounce and that on the trivial 
configuration can be thus written as
\beq
S_B=\pi \, (T_2-2 \, T_1) \, r_c^2 + 4 \pi \, T_1 \, \int_{r_c}^R \, r \, dr \, \left ( \sqrt{1+z^{\prime\, 2}}-1 \right )  ~,
\label{sbint}
\eeq
where $z(r)$ stands for the vertical profile of one of the two world sheets (for definiteness we consider the upper one) and the contribution of the other simply doubles the coefficient in front of the integral in Eq.~(\ref{sbint}).

\begin{figure}[h!t]
\epsfxsize=9cm
\centerline{\epsfbox{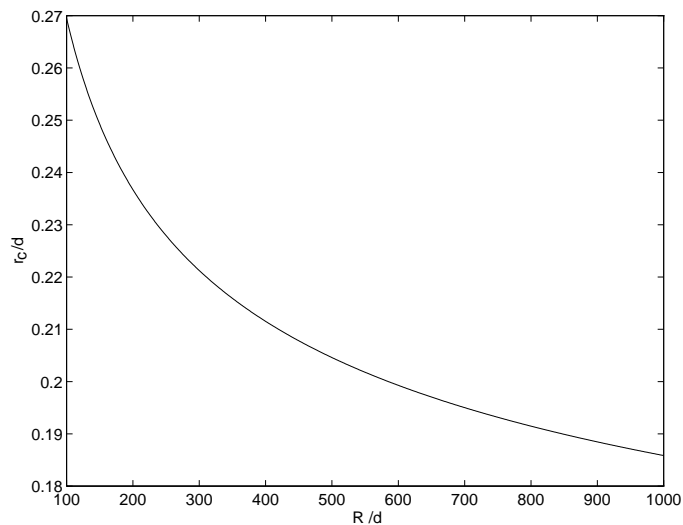}}
\caption{{\footnotesize  The plot of $r_c/d$ vs. $R/d$ at $\smallfrac{T_2}{2T_1}=0.95$ (i.e. $\delta \approx 0.31 $, see Eq.~(\ref{del})).}}
\label{five}
\end{figure}

We find that the simplest way to analyze the solution for $z(r)$ is using an ``integral of motion", which follows from the symmetry under $z$ translation, 
which we call $r_0$:
\beq
{r \, z' \over \sqrt {1+z^{\prime\,2}}} = r_0 \ .
\label{r0}
\eeq
The left-hand side is independent on $z$,
which, in fact, tells us that the vertical component of the capillarity force acting on any horizontal section of the film is constant. The relation between the constant $r_0$ and the radius $r_c$ of the bounce is found from the condition of equilibrium of the boundary of the disk, where the string $T_2$ bifurcates into two strings $T_1$. This condition is that the net horizontal force at the boundary vanishes,
\beq
\left . {2 \, T_1  \over \sqrt{1+z^{\prime\,2}}}
\rule{0mm}{8mm}\,\,
\right |_{r=r_c}=T_2~.
\label{hforce}
\eeq
After setting $r=r_c$ in Eq.~(\ref{r0}) and eliminating $z'\,|_{r=r_c}$ from Eqs. (\ref{r0}) and (\ref{hforce}),
one readily finds the following relation:
\beq
r_0=r_c \, \sqrt{1-{T_2^2 \over 4 T_1^2}}~.
\label{r0rc}
\eeq
The solution to the equation of motion (\ref{r0}) satisfying the boundary condition $z(R)=d/2$ at $R \gg r_c$ 
has the form\,\footnote{To be more exact, in Eq.~(\ref{r0rc})  $z(R) = d/2$ up to terms $O\left(\frac{r_0^2}
{R^2}\,\frac{d}{r_0}\right)$. As we will see shortly, roughly speaking, 
$r_0 \sim d/\left[2\ln (R/d)\right]$. Hence, up to logarithms, the relative error is $O\left(\frac{d^2}{R^2}
\right)$ and is negligible due to condition (\ref{limitation}).}
\beq
z=r_0 \, \ln {r+\sqrt{r^2 - r_0^2} \over 2 R} + \, \frac{d}{2}~.
\label{zsol}
\eeq
The parameter  $r_c$ can be determined  from the condition $z(r_c)=0$ in terms of $d$ and $R$. 
To this end we substitute Eq.~(\ref{r0rc}) in 
\beq
\ln \frac{2R}{\,r_c+\sqrt{r_c^2-r_0^2}}=\frac{d}{2r_0}\,,
\eeq
which can be solved numerically. Figure \ref{five} presents $r_c/d$ as a function
of $R/d$ at a representative value of $T_2/(2T_1)=0.95$.
(A matching with Eq.~(\ref{dellimit}) below starts emerging at the right edge of the plot.)

Given the solution (\ref{zsol}) it is not difficult to
find the action (\ref{sbint}). It turns out that
the action $S_B$ has the simplest form being expressed
in terms of $r_0$ rather than $r_c$, namely,
\beq
S_B=\pi \, T_1 \, r_0 \, (d - r_0 )~.
\label{sbr0}
\eeq
Note that the boundary condition at $r=r_c$ written in terms of $r_0$ reads 
\beq
r_0 \, \ln \left ( {r_0 \over 2 R} \, \sqrt{{2T_1+T_2 \over 2 T_1-T_2}} \right ) = -{d \over 2}~.
\label{r0bc}
\eeq

Equations (\ref{sbr0}) and (\ref{r0bc}) provide the solution for the exponential factor 
in the probability of the merger of two strings for arbitrary ratio of the tensions 
$T_2 / (2T_1)$. In particular at $T_2=0$ the problem is equivalent to that of spontaneous reconnection of two parallel strings\cite{Tong}. From (\ref{r0rc}) we learn that in this limit 
\beq
r_0=r_c\,,
\eeq
and the bounce configuration is described by a configuration
discussed in~\cite{Gorsky}. In the present
paper we will focus on the opposite limit in which
the binding of the strings is parametrically small, i.e. 
\beq
2T_1-T_2 \ll T_1\,. 
\eeq
It is convenient to introduce a dimensionless small parameter $\delta$ for the binding,
\beq
\delta= \sqrt{1-{T_2^2 \over 4 T_1^2}} ~.
\label{del}
\eeq
One can readily verify that in the limit 
$\delta \ll 1$
the gradient of the deviation of the string profile
from a flat string is small, and the equations for the profile of 
the string world sheet in the bounce configuration 
can be {\em linearized}. This allows one to consider the fusion problem in a
more physical
strip geometry, i.e. with the strip  boundary conditions.
This can be done both
for parallel strings and slightly nonparallel ones.

\subsection{Linearizing the problem in the weak binding limit}
\label{linear}

In the linearized approximation (valid if $|\vec\nabla z |\ll 1$)
the classical equation of motion for the string profile is
\beq
\Delta z=0\,.
\label{laplac}
\eeq
If $\delta\ll 1$ the above condition is met. It is not difficult to solve Eq.~(\ref{laplac}) with the round
boundary condition. Alternatively, one can expand
the full nonlinear solution in the disk geometry.  One finds in the leading logarithmic in $R$ approximation
at $\delta \ll 1$ 
\beq
r_0 = r_c \, \delta~~ {\rm and}~~ r_0 \approx {d \over 2 \, \ln(R \delta/d)}~.
\label{dellimit}
\eeq
Then the action $S_B$ can be readily derived from  Eq.~(\ref{sbr0}), namely,
\beq
S_B \approx \pi \, T_1 \, {d^2 \over 2 \, \ln(R \delta/d)} \, \left \{ 1 + O \left [ {1 \over \ln (R/d)} \right ] \right \}~.
\label{sbdel}
\eeq
Please, remember that the exponent determining the decay rate is $e^{-S_B}$. The condition (\ref{ishco}) justifies the quasiclassical approximation. 

It is interesting to note that the bounce action is mainly determined 
by the tension $T$ and the distance $d$, rather than by the binding parameter $\delta$.
The formula (\ref{sbdel}) also tells us that the bounce action becomes small, 
and the semiclassical treatment becomes inapplicable, for exponentially long 
strings. However it is clear from the overall proportionality of the 
fusion rate to the string length  that for such long strings the 
probability of fusion becomes of order one. It can also be readily verified 
that introducing of a small mass $\mu$ for the three-string junction, 
neglected throughout this paper, does not change the infrared dependence of 
the bounce action. Indeed, the $\mu$-induced contribution to the action is 
\beq
\Delta S_B = 2 \pi \, \mu \, r_c \approx {\pi \, \mu \, d \over \delta \, \ln(R \delta/d)}~,
\label{mudsb}
\eeq
which has the same logarithmic behavior at large $R$ as $S_B$ in Eq.~(\ref{sbdel}). Thus, the condition under which the junction mass can be neglected does not depend on the string length and reduces to
\beq
\mu \ll d \, \delta T_1~.
\label{mucond}
\eeq

\subsection{Strip boundary condition in linear approximation}
\label{sbcla}

In the linear approximation in which
the equation of motion for $z(x,\tau)$ reduces to the two-dimensional Laplace equation
(\ref{laplac}),
the solution can be constructed as a real (or imaginary) part of a holomorphic function of the complex variable \beq
w=x+i \, \tau\,.
\label{www}
\eeq
Using this construction and the analogy with  two-dimensional electrostatics (the so-called
image charges method, see Appendix~B), one can readily find the solution for the 
strip $ -L/2 \le x \le L/2$ with the boundary conditions $z(\pm L/2)=d/2$. 
For the bounce centered at $x=l$ and $\tau=0$ the solution has the form
\beq
z(x, \tau)= r_0 \, {\rm Re} \left [ \ln \left({\sin {\pi (w- l) \over 2 L} \over \cos {\pi (w+l) \over 2 L}} 
\right)\right ]+\, {d \over 2}~,
\label{zstrip}
\eeq
with the constant $r_0$ being determined by the condition of equilibrium of the bifurcation boundary, corresponding to $z=0$,
similarly to Eq.~(\ref{r0rc}). Clearly, in the strip geometry the $O(2)$ symmetry is lost and 
the world sheet boundary for the string $T_2$ is no longer a disk. However for large $L$ and for the bounce center not too close to the strip edge, the exact solution (\ref{zstrip}) can be approximated by a logarithmic one,
\beq
z \approx r_0 \, {\rm Re} \left [ \ln  {w- l \over L} \right ] + {\rm const}~.
\label{zsapp}
\eeq
It corresponds to an
approximately circular bifurcation boundary with the radius $r_c$ related to the parameter $r_0$ as 
in Eq.~(\ref{dellimit}). The applicability conditions for  
this approximation are that $\ell$ is not parametrically close to $L/2$ and also $L \gg r_c$.
The bounce action $S_B$ on such configuration in the logarithmic in $L$ approximation coincides with 
that in Eq.~(\ref{sbdel}) with $R$ being replaced by $L$,
\beqn
S_B &\approx &\pi \, T_1 \, {d^2 \over 2 \, \ln(L \delta/d)} \, \left \{ 1 + O \left [ {1 \over \ln (L/d)} \right ] \right \}\,,
\nonumber\\[4mm]
\delta &\approx& \sqrt{\frac{2T_1-T_2}{T_1}}\,.
\label{sbdelp}
\eeqn

\section{Nonparallel strings}
\label{nonpar}

\begin{figure}[h!t]
\epsfxsize=7cm
\centerline{\epsfbox{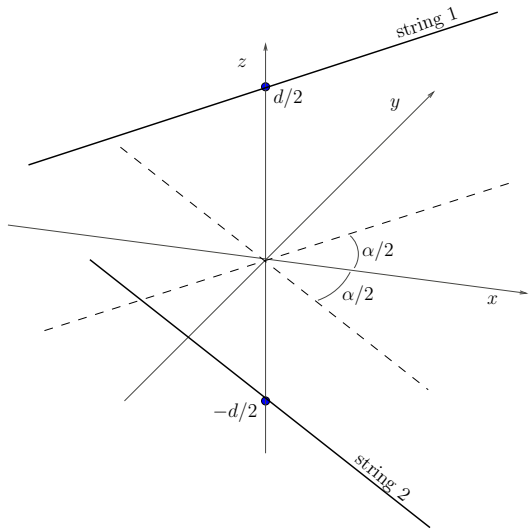}}
\caption{{\footnotesize  Geometry for nonparallel strings.  }}
\label{sghemb}
\end{figure}

When the number of the space-time dimensions is $3+1$ (or more), it is possible to 
have nonintersecting and nonparallel strings.  
Then geometry of the problem can be characterized by two parameters:
the minimal distance $d$, and an angle $\alpha$, which we will assume to be small (Fig.~\ref{sghemb}). 
We chose the spatial axes in the following way, the $z$ axis runs along the common perpendicular to the strings, and the origin of the coordinates is placed in the middle of the segment of this perpendicular connecting the strings. 
The $x$ axis runs along the bisector of the angle between the projection of the strings on 
the $\{x,y\}$ plane, so that  the strings  are parametrized as ($\alpha \ll 1$)
\beqn
z&=&
d/2\,,\quad  y = \tan \frac{ \alpha\, x}{2}\,,
\nonumber\\[1mm]
z&=&
-d/2 \,,\quad  y = -\tan \frac{ \alpha\, x}{2}\,.
\eeqn 
We will refer to this configuration as ``twisted strings."
In choosing the infrared regularization we  aim at  logarithmic accuracy of the bounce 
action, so that it is sufficient to consider the axially symmetric geometry of the world sheet with 
a large radius $R$, i.e. the round boundary conditions (Fig.~\ref{boundary}). 
Namely, the following constraints will be imposed:
\beq
z(R) = \pm\, \frac{d}{2}
\eeq
for  two strings under consideration, while
the boundary conditions for $y$ are 
\beq
y(x,\tau) \Big|_{r=R}= \pm {\alpha \over 2} \, x~.
\label{bcy}
\eeq
In the linearized approximation the equations of motion for the 
orthogonal deviations of the string $z(x,\tau)$ and $y(x, \tau)$ 
are independent from one another.
If $\alpha$ is small,  ``twisting" in the variable $y$ can be considered 
as a perturbation over the solution  $z(x,\,\tau)$ for parallel strings
(Sect.~\ref{strings}). 
It is clear from the symmetry of the problem that $y$ must
identically vanish in the central part of the bounce, i.e. at  $r \le r_c$ one has $y=0$. 
Then between the circles at $r=r_c$ and $r=R$ the function $y$ is harmonic and changes from $y(r_c)=0$ to the values prescribed by the boundary conditions (\ref{bcy}). Invoking the well-known central harmonics for the two-dimensional Laplace operator, we find the explicit form of the profile $y(x,\tau)$ for the upper and the lower strings,
\beq
y= \pm {\alpha \over 2} \, {x \over 1-r_c^2/R^2} \, \left ( 1- {r_c^2 \over r^2} \right )~.
\label{yshape}
\eeq


Thus, the twist in the $y$ direction results in an additional contribution to the (linearized)
bounce action which takes the form
\beq
\delta_y S_B = {7 \pi \, \alpha^2 \over 16 } \, T_1 \, r_c^2~.
\label{dysb}
\eeq
Using the relations (\ref{dellimit}) for $r_c$ one finds that the fusion probability 
for twisted strings $\Gamma(\alpha)$ is reduced in comparison with the parallel strings,
\beq
\Gamma(\alpha) = \Gamma(0) \, \exp \left [ - {7 \pi \, \alpha^2 \over 64 \, \delta^2 } \, T_1 \, {d^2 \over \ln^2(R \delta/d)} \right ]~.
\label{ares}
\eeq
Here $\Gamma (0)$ is the merger probability for the parallel strings.
It can be noted that, although the contribution to the exponential factor
associated with the twist  is of a higher order in  $1/\ln(Rd^{-1})$ than
that in $\Gamma (0)$, it has a nontrivial singular dependence on $\delta$. The latter dependence implies that the string merger at weak binding takes place only if the angle between them is also small, i.e. at $\alpha < \alpha_{\rm max}$ where
\beq
\alpha_{\rm max} \sim \delta \,\, {\ln(R \delta/d) \over \sqrt{T_1} \, d}~.
\label{amax}
\eeq

\section{Fusion of parallel domain walls}
\label{walls}

We will analyze the domain wall fusion at $2T_1 -T_2\ll T_1$,
when the linearized approximation is applicable.
In contradistinction with
the string problem,  in the domain wall problem we do not need any infrared regularization,
since the solution of the three-dimensional Laplace equation falls off as $1/r$ rather
than logarithmically. 
The coordinates  that parametrize the Euclidean $3$-brane are $x,\,y,\,\tau$, while
the walls are given by the height functions $z=f(x,\,y,\,\tau)$.
The boundary conditions are set at infinite $r = (x^2=y^2+\tau^2)^{1/2}$,
\beq
z (r) \to \pm \, \frac{d}{2} \,\,\, \mbox{at}\,\,\, r\to\infty .
\eeq

The solutions of the linearized equation $\Delta z =0$ for the top and bottom walls are
\bea
z_1 (x,\,y,\,\tau) &=& -  \frac{A}{r}  + \frac{d}{2} \ , \nonumber \\[3mm] 
z_2(x,\,y,\,\tau) &=&   \frac{A}{r} - \frac{d}{2} \label{sol} \ ,
\label{zlina}
\eea
where  the bounce is assumed to be centered at the origin.
The two $T_1$ branes meet at $r= r_c $ where 
\beq
r_c = \frac{2A}{ d}\,.
\label{rsubc}
\eeq
It is obvious that $r_c$ is the radius of the word volume of the composite
wall (i.e. the world volume radius of the $T_2$ brane configuration) at the moment it leaves 
Euclidean and enters the Minkowski space ($\tau =0$).

\begin{figure}[h!t]
\epsfxsize=12cm
\centerline{\epsfbox{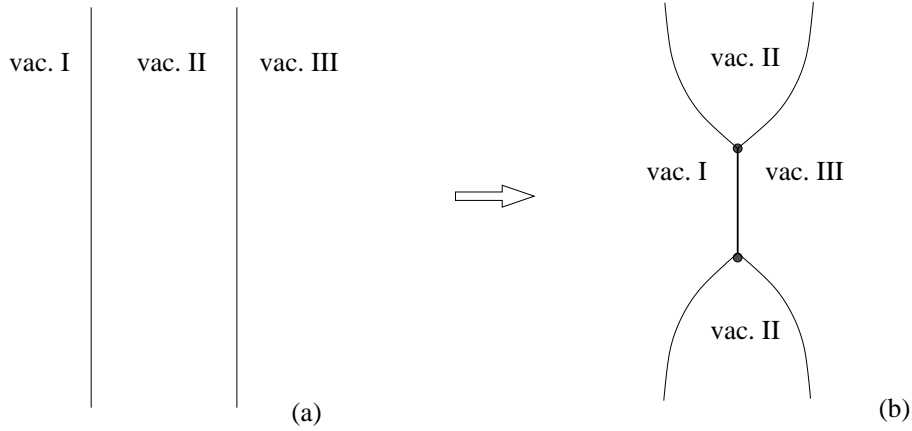}}
\caption{{\footnotesize      Geometry of the domain wall fusion   }}
\label{wallsfusion}
\end{figure}

The total Euclidean action is the sum of two contributions
\beqn
S
&=&
\left(T_2 -2T_1\right) \, \frac{4\pi }{3}\, r_c^3  +  2 T_1 \int_{r_c}^{\infty} \,4\pi r^2 dr \,\,
\frac{ z^{\prime\,2} }{2}
\nonumber\\[3mm]
&=&
- \left(2T_1 -T_2\right) \, \frac{4\pi }{3}\, r_c^3 +T_1\,\pi\,r_c\, d^2\,,
\label{regac}
\eeqn
where the first one comes from the composite wall in the middle, while the second from two tails of elementary walls.
The action (\ref{regac}) is regularized: we subtracted the contribution
of two parallel undistorted walls (Fig.~\ref{wallsfusion}$a$). In deriving this action we used Eq.~(\ref{rsubc}).

Please, note that the signs are opposite. The first term is negative, and dominant at large $r_c$. The second is positive and dominant at small $r_c$. The bounce solution which is at the tip of hill
can be obtained by extremizing Eq.~(\ref{regac}) with respect to $r_c$,
\beqn
r_c &=&   \frac{d}{2}\, \sqrt{\frac{T_1}{2T_1 - T_2}} \, ,  
\nonumber\\[3mm]
S_{B} &=&  \frac{\pi}{3}\,\, { T_1 d^3}\, \sqrt{\frac{T_1}{2T_1 - T_2}} \, .
\label{rcsb}
\eeqn
The probability of the wall fusion per unit time and unit area is proportional to
$e^{-S_B}$.

Now we can   check that the linearization approximation is valid. 
The necessary condition is $|z'|\ll 1$ which is equivalent to $A /r_c^2 \ll 1$. Equations
(\ref{rsubc}) and (\ref{rcsb}) imply
\beq
\frac{A}{r_c^2} \sim \frac{d}{r_c} \sim \sqrt{\frac{ 2T_1 - T_2}{T_1}}\,.
\eeq
This condition is met at weak binding, i.e.
\beq
\frac{2T_1 -T_2}{T_1} \ll 1 \, .
\label{wb}
\eeq
Note that it does not depend on the inter wall distance $d$. However, the 
distance must be much larger than the wall thickness. It must be large 
enough to ensure $S_B\gg 1$. In particular, the choice 
$T_1\,d^3 \gg 1$ does the job. Another condition that was assumed 
in the consideration above is that the tension of the three-wall junction 
(closed circles
on Fig.~\ref{wallsfusion}$b$) is negligible, so that the junction 
contribution to the action can be ignored.

\section{Parallel walls at strong binding}
\label{pwlb}

If Eq.~(\ref{wb}) is not satisfied, the wall binding is strong, and the bounce action must be treated
beyond the linear approximation,
\beq
\label{tru}
S_B = \frac{4\pi}{3} \,  \left( T_2- 2T_1 \right)\,r_c^3
+ 2 T_1 \int_{r_c}^{\infty}4 \pi \, r^2 dr\, \left( \sqrt{ 1 + z^{\prime\,2} } - 1 \right) \ .
\eeq
It is not difficult to see
that there exists  an ``integral of motion" analogous to (\ref{r0}),
\beq
{r ^2\, z' \over \sqrt {1+z^{\prime\,2}}} = r_0^2~.
\label{r0p}
\eeq
The solution for the wall bounce must be such that the left-hand side of
(\ref{r0p}) is $r$ independent. We use the notation $r_0^2$
for this constant. Then the classical equation of motion reduces to
\beq
\frac{dz}{dr} =\pm\,{\frac{r_0^2}{\sqrt{r^4-r_0^4}}}\,,
\label{41}
\eeq
implying the following relation between the parameters $r_0$,   $r_c$ and $d$:
\beq
r_0^2 \, \int_{r_c}^\infty {dr \over \sqrt{r^4-r_0^4}} = {d \over 2}~.
\label{r0rcd1}
\eeq
Another relation between $r_0$ and $r_c$ is the equilibrium condition coinciding with Eq.~(\ref{hforce}). Considering the expression (\ref{r0p}) for the integral of motion in the wall case, one readily finds from this condition that
\beq
r_0^4 = r_c^4 \, \left ( 1- {T_2^2 \over 4 \, T_1^2} \right )~.
\label{r0rc2}
\eeq
Invoking Eqs.~ (\ref{r0p}) and (\ref{41}), the bounce action in Eq.~(\ref{tru}) can be transform\-ed to 
\beqn
S_B 
&=&
- \left(2T_1 - T_2\right)\frac{4\pi}{3} \, r_c^3 + 8 \pi \, T_1 \, \int_{r_c}^\infty \, dr\,\left ( {r^4 \over \sqrt{r^4-r_0^4}} - r^2 \right )
\nonumber \\[3mm]
&\equiv&
\frac{4\pi}{3} \, T_2\,r_c^3 +  \frac{8\pi}{3} \, T_1\,  \int_{r_c}^\infty \, dr\,\frac{r_0^4}{\sqrt{r^4-r_0^4}}
\nonumber \\[3mm]
&-&
\frac{8\pi}{3} \, T_1\, \left[ r_c^3 -\,\int_{r_c}^\infty\, dr\,
\left(
\frac{3r^4-r_0^4}{\sqrt{r^4-r_0^4}} - 3r^2
\right)
\right]\,.
\eeqn
The second term in the first line is an elliptic integral which we will
replace by its value mandated by the condition (\ref{r0rcd1}).
The expression in the square brackets in the second line
reduces to a combination of elementary functions.
Indeed,
\beq
r_c^3 -\,\int_{r_c}^\infty\, dr\,
\left(
\frac{3r^4-r_0^4}{\sqrt{r^4-r_0^4}} - 3r^2
\right)
= r_c \,\sqrt{r^4_c-r_0^4}\,. 
\label{psb22}
\eeq
This can be directly verified by differentiating  both sides in Eq.~(\ref{psb22}) over $r_c$.
As a result, the bounce action takes the form
\beq
S_B=
{ 4 \pi \over 3} \, \left ( T_2 \, r_c^3 - 2 \, T_1 \, r_c \,\sqrt{r_c^4-r_0^4} +  T_1 \, r_0^2 \, d \right )\,.
\label{sb22}
\eeq
After substituting in the latter expression the relation (\ref{r0rc2}) one readily finds that the first two terms in parentheses cancel, and one is left with a simple formula for the bounce action in terms of $r_0$ and $d$,
\beq
S_B= {4 \pi \over 3}\, T_1 \, r_0^2 \, d~. 
\label{sbr0d2}
\eeq 
In order to express the parameter $r_0$ in terms of the separation distance $d$ and the wall tensions $T_1$ and $T_2$ one needs to solve the transcendental equation (\ref{r0rcd1}) which results in an elliptic function. In the limit of weak coupling one recovers the results of the previous section, while in the extreme strong coupling limit, $T_2 \to 0$, one has $r_c=r_0$ and we find
$$ r_0 = {\Gamma \left ({3 \over 4} \right ) \over 2 \, \sqrt{\pi} \, \Gamma \left ({5 \over 4} \right ) } \, d =0.381 \ldots d ~,  $$
so that the bounce action can be written as
\beq
S_B \Big|_{T_2 \to 0} = 
{1 \over 3} \left [ {\Gamma \left ({3 \over 4} \right ) \over  \Gamma \left ({5 \over 4} \right ) } \right ]^2\, T_1 \, d^3 = 0.609 \ldots T_1 \, d^3~.
\label{sbt20}
\eeq

Thus, an estimate $S_B = {\rm const}.\,\pi\, T_1\, d^3\, \delta^{-1}$ works in the entire range of variation
of $T_2$ in the problem of the domain wall fusion: from $T_2=2T_1$ (where $\delta \to 0$) down to $T_2=0$.

\section{An instructive example}
\label{aie}

In this section we will consider a particular example in which weakly bound
domain walls naturally appear. Supersymmetric Yang--Mills theory supports \cite{bpsdw}
critical domain walls whose tensions are exactly known \cite{bpsdw,bpsdwp}. Namely, the tension of the $k$-wall
is given by the formula
\beq
T_k = N^2\, \Lambda^3 \,\sin\left(\frac{\pi k}{N}\right)
\eeq
for SU$(N)$ gauge group. Here $\Lambda $ is a dynamical scale parameter. 
The tension of the composite $k$ walls  is less than $k$ times the elementary wall tension
(i.e. the $k=1$ wall).

Hence, at large $N$ we have
\beq
2T_1 - T_2 = \frac{\pi^3}{N}\, \Lambda^3\,,\qquad T_1 = {\pi}{N}\,\Lambda^3\,.
\eeq
Invoking Eq.~(\ref{rcsb}) we conclude that the probability of two parallel wall fusion (per unit time per unit 
area) is proportional to
\beq
\Gamma \sim \exp\left[-\frac{\pi}{3}\,N^2 \left(\Lambda\, d\right)^3
\right]\,.
\eeq

\section{Conclusions}
\label{conq}

In this paper we have considered a generic problem of restructuring (fusion)
of solitonic objects, due to binding energy. 
We paid particular attention to formulating, and solving, the problem in a generic way, without any reference to  particular underlying theories, or mechanisms responsible for the soliton existence and binding (i.e. no reference to microscopic physics). 
In this way our results can be applied in every instance in which our assumptions (summarized at the very end of  Introduction) are met.
Although we have focused on strings and walls in $3+1$ dimensions, our results can be easily generalized to higher-dimensional branes and/or higher space-time dimensions. 
For example, if we consider two $p$ branes in $(p+2)$-dimensional space-time,
the fusion probability per unite volume and unit time is
\beq
\Gamma_p \sim\exp\left\{ - C_{p+1} \, 2^{-p}\, (p-1)^p\,\left(\frac{T_1}{2T_1-T_2}\right)^{(p-1)/2}\, \left(T_1\, d^{p+1}\right)
\right\}\,,
\eeq
where $C_\ell$ is the volume coefficient,
\beq
C_\ell =\frac{\pi^{\ell/2}}{\Gamma\left(\smallfrac{\ell}{2}+1\right)}\,.
\eeq
The problem for strings 
in $3+1$ dimensions has a peculiar infrared behavior, and requires a regularization. For domain walls, or higher-dimensional branes, no infrared regularization is needed.

The study and computation of the fluctuations around the bounce solution is left  for future work  This problem is essential in order to compute the coefficient ${\cal C}$ in Eq.~(\ref{gamsb}). In some circumstances, it is just a numerical coefficient, which have little impact on the physical behavior.
In other cases it can be of crucial importance.
Assume we want to study the problem of infinite, parallel strings.
We thus take the problem considered in Sects. \ref{round} and \ref{linear}, and send the cutoff $R \to \infty$.  
In this case $r_c$ and $r_0$ both go to zero, and  the bounce action 
formally vanishes.  The center of the bounce is located at the center of the circle of radius $R$. Translation of the center becomes approximately a flat direction, as $R \to \infty$. This is a usual divergence that is absorbed in the 
volume dependence,  so that in fact we compute the decay probability per unit time and  unit length. 
But there is also another (nearly) flat direction -- the one due to a rescaling $(x,\tau) \to \lambda (x,\tau)$.
This is peculiar to the case of strings. In the weak binding limit, the right-hand side of (\ref{sbint}) becomes 
(nearly) scale invariant. 
This extra (approximate) zero mode must be properly treated  in calculating  the factor ${\cal C}$.

Is worth to mention a particular example of binding energy between strings. It arises in type IIB string theory where $(p,q)$ strings are bound states of $p$ F$1$-strings and $q$ D$1$-strings.
Networks of $(p,q)$-strings have been studied recently, since they naturally arise at the end of some string theory inflation scenarios \cite{Tye:2005fn}. Tunneling effects may be relevant for the evolution of these strings networks.

Finally, we would like to mention that work on the D-brane fusion 
was carried out (e.g. \cite{dop})
in string theory, in a setting specific to string theory.

\section*{Acknowledgments}
We would like to thank A. Armoni and A.~Vainshtein for usefull discussions.
This work  is supported in part by DOE grant DE-FG02-94ER408.

\section*{Appendix A: Connection with field theory}
\label{b}

\renewcommand{\theequation}{A.\arabic{equation}}
\setcounter{equation}{0}

\begin{figure}[h!t]
\epsfxsize=10.5cm
\centerline{\epsfbox{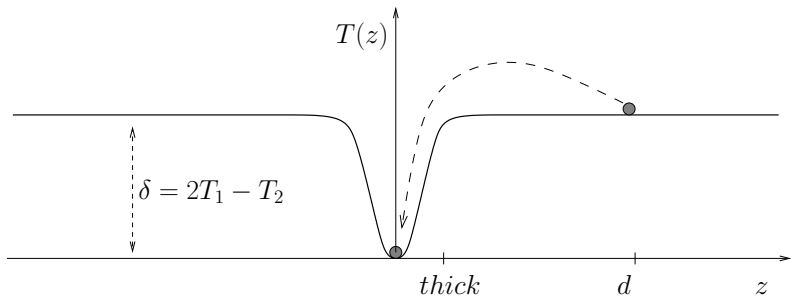}}
\caption{{\footnotesize Potential as a function of the distance $z$.  }}
\label{falseplateaux}
\end{figure}

In the case of weak binding, and parallel branes, the problem can be recast in a simple field-theoretic 
formulation. Let us discuss it for the specific case of parallel strings.
The ``true'' action is the Nambu-Goto one,
with a tension that depends upon the distance between the two strings.
For small perpendicular fluctuations $z$, i.e. at weak binding, we can use $x$  to parametrize 
the space coordinate on the world sheets of the strings, and  the following action ensues:
\beqn
S  &=& - T(z) \int dx d\tau \, \sqrt{1- \partial_{\mu}  z \,\partial^{\mu} z } \nonumber \\[3mm]
& = &  \int dx d\tau \  T(z) \left( -1 + \frac{1}{2} \partial_{\mu} z\, \partial^{\mu} z  + \dots \right),
\eeqn
where $z$ is the distance between two strings, and $T(z)$ is the ``combined tension" as a function of the distance. 
If we define a scalar field $\phi$,
\beq
\phi = z\sqrt{T(z)} \ ,
\eeq
assuming that the $z$ dependence of $T$ is adiabatically slow, we get the action in the form
\beq
S = \int dx d\tau \  \left(  \frac{1}{2} \,\partial_{\mu}  \phi \,\partial^{\mu} \phi  -V(\phi) \right)  ,
\eeq
with the  canonically normalized kinetic term
and
\beq
V(\phi) = T(z(\phi)) \, .
\eeq
The shape of the effective potential $V(\phi)$ is very similar to $T(z)$, although the change of variables 
from $z$ to $\phi$ changes the functional dependence in the vicinity of the minimum at $z=0$, Fig.~\ref{falseplateaux}. 
This effective potentail is of a ``false plateau'' type. It is flat almost everywere, apart from a small domain near zero, where it drops off by $2T_1-T_2$.

Details of the potential shape near the minimum depend on microscopic physics that is responsible for 
the binding energy. 
The results presented in this paper do depend  on these details.  What was crucial was
the fact  that the effective potential is essentially constant and flat everywhere,  and drops to zero at a very short distance
(the string/wall thickness) from the origin.

The flat plateaux means that we have a classical moduli space of vacua, which, in turn, 
corresponds to the fact that there are no long-range forces between the two strings (or walls).
The tunneling is  similar to the conventional false vacuum decay studied for metastable  potentials \cite{four}. A key
technical difference is that a barrier is present only in the form of kinetic energy.

We can add the following regulator to the potential:
\beq
V_{\epsilon} = \epsilon z^2  (z - d)^2\,,\qquad \epsilon \to 0\,.
\eeq
The total potential $T(z) + V_{\epsilon}(z)$ mantains $z=0$ as a true vacuum. 
The plateau is lifted, and $z=d$ is a metastable vacuum. In this formulation, the
decay probability is determined by the conventional bounce
of the type relevant to the false vacuum decay.
At $\epsilon \to \infty$  the thin wall approximation is valid. On the other hand, at   $\epsilon\to 0$
(the case we are interested in) we find ourselves completely outside of the thin wall approximation, albeit,
as we see in the bulk of the paper, a bounce-like solution exists and can be explicitly found.

The bounce becomes exceedingly shallower as we decrease $\epsilon$.
The number of space-time dimensions is crucial here. 
In $(1+1)$-dimensional theory,  the solution asymptotically is logarithmic, and the boundary
condition $z_1-z_2= d$ can only be imposed at a finite distance.
In three or more dimensions the fall-off is power-like, and the boundary condition
can be imposed at infinity.

\section*{Appendix B: Holomorphic potentials}
\label{a}

\renewcommand{\theequation}{B.\arabic{equation}}
\setcounter{equation}{0}

\begin{figure}[h!t]
\epsfxsize=11cm
\centerline{\epsfbox{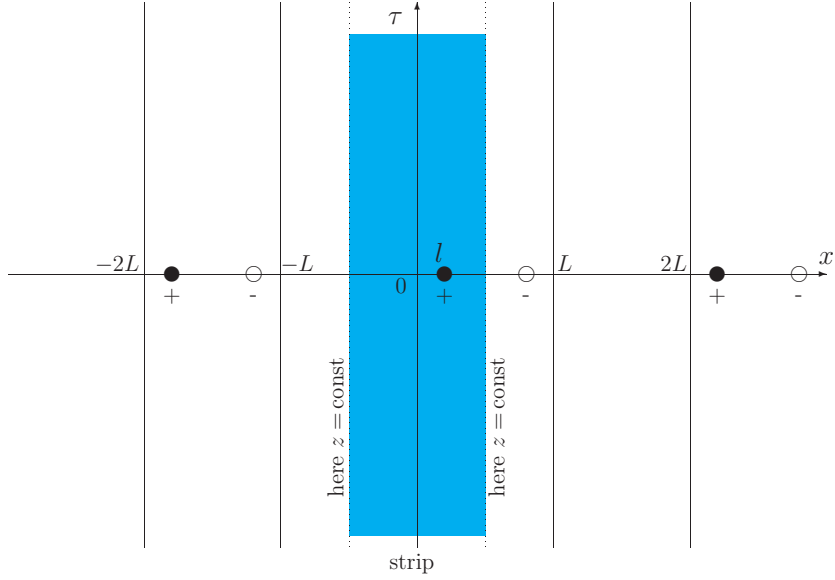}}
\caption{{\footnotesize   The system of mirror `charges'  
relevant to deriving the solution in Sect.~\ref{sbcla},
with strip boundary conditions. The shaded strip is the physical world sheet for the strings and all the sidebands are its `images'. The filled circles stand for positive charges and the open circles are for the negative `images'.}}
\label{holopot}
\end{figure}

A good way to  directly derive Eq.~(\ref{zstrip}) in Sect. \ref{sbcla} is 
provided by analogy   with  two-dimensional electrostatics through
the well-known in electrostatics  image charges method. To this end we
use the trick of putting auxiliary ``mirror" charges
as shown in Fig. \ref{holopot}.
We  extend the strip to the entire complex $w=x+i\tau$ plane, 
put the original unit charge at $w=l$, 
and then add a series of `-1' mirror charges at $w_k^{-} = (2k +1) \, L -l$, and a series of `+1' mirror charges at $w_k^{+} = 2k  \, L +l$, with $k$ being any integer. (In fact $k=0$ in the latter case corresponds to the original charge whose potential is being calculated with the strip boundary conditions.)  
In this way we certainly satisfy, due to the symmetry of the system of mirror charges, the boundary conditions that $z$ must vanish at Re$\,w = \pm L/2$.  
We then can find the solution for the harmonic function $z(x,\tau)$ in terms of the real part of a holomorphic potential $\Phi(w)$ produced by the constructed system of charges: $z=C_1 \, {\rm Re} \Phi(w) +C_2$ with $C_1$ and $C_2$ being arbitrary constants.
The holomorphic potential from the considered system of charges is given by
\bea
\Phi(w) &=& \sum_{k \in \Z}  \ln (w -w_k^{+}) - \sum_{k \in \Z}  \ln (w -w_k^{-}) \nonumber \\
&=&  \ln \prod_{k \in \Z} (w-w_k^+) -  \ln \prod_{k \in \Z} (w-w_k^-)\,. 
\label{holp}
\eea
Using the Euler's formula representing the $\sin$ function as a product, one can explicitly find the products in Eq.(\ref{holp}) up to inessential (although infinite) multiplicative constant:  
\beqn
\prod_{k \in \Z} (w-w_k^+) = {\rm const} \cdot \sin \frac{\pi (w-l) }{2L} \,,
\nonumber\\ [3mm]
\prod_{k \in \Z} (w-w_k^-) = {\rm const} \cdot \cos \frac{\pi (w+l) }{2L} \,.
\eeqn
From here  the harmonic function (\ref{zstrip}) ensues.
It is certainly easy to check, {\it a posteriori}, that the function in Eq.(\ref{zstrip}) satisfies the necessary boundary conditions in the discussed problem.


\end{document}